\PassOptionsToPackage{hyphens}{url}
\documentclass{article}

\usepackage[preprint]{neurips_2024}

\usepackage[utf8]{inputenc} 
\usepackage[T1]{fontenc}    
\usepackage{hyperref}       
\usepackage{microtype}      
\usepackage{xcolor}         
\usepackage{url}            

\title{Recommended Practices for NPOV Research on Wikipedia}

\author{
  Isaac Johnson\\
  Wikimedia Foundation\\
  \texttt{isaac@wikimedia.org} \\
  \And
  Yu-Ming Liou\\
  Wikimedia Foundation\\
  \texttt{yliou@wikimedia.org} \\
  \And
  Jacob Rogers\\
  Wikimedia Foundation\\
  \texttt{jrogers@wikimedia.org} \\
  \And
  Aaron Shaw\\
  Northwestern University\\
  \texttt{aaronshaw@northwestern.edu} \\
  \And
  Leila Zia\\
  Wikimedia Foundation\\
  \texttt{leila@wikimedia.org} \\
}

\begin{document}

\maketitle

\begin{abstract}
  Writing Wikipedia with a neutral point of view is one of the five pillars of Wikipedia.\footnote{\url{https://en.wikipedia.org/wiki/Wikipedia:Five_pillars}} Although the topic is core to Wikipedia, it is relatively understudied considering hundreds of research studies are published annually about the project.~\cite{asikin2025research} We hypothesize that part of the reason for the low research activity on the topic is that Wikipedia’s definition of neutrality and its importance are not well understood within the research community. Neutrality is also an inherently challenging and contested concept. Our aim with this paper is to accelerate high quality research in this space that can help Wikipedia communities continue to improve their work in writing the encyclopedia. We do this by helping researchers to learn what Neutral Point of View means in the context of Wikipedia, identifying some common challenges with studying NPOV and how to navigate them, and offering guidance on how researchers can communicate the results of their work for increased impact on the ground for the benefit of Wikipedia.
\end{abstract}

\section{Introduction}
Wikipedia stands as the largest encyclopedia in history and one of the internet's most widely used information resources. Every month, hundreds of thousands of volunteers collaboratively build and refine over 300 language editions, striving to deliberate and document the \textit{sum of encyclopedic human knowledge}. Wikipedia articles are viewed over 15 billion times per month directly on the platform, with billions more accessing its freely available content indirectly through third-party applications, services, and widely used technologies. Wikipedia’s open approach to content availability has also made it a valuable source of data for other services and projects, including natural language processing models and large language models~\cite{johnson2024wikimedia}.

This vast reach and persistent success are not accidental; rather, they are the outcome of dedicated work by volunteers across the globe. These contributors are engaged not only in the research, writing, and improvement of articles, but also in the crucial processes of deliberation, discussion, and consensus building—a collaborative effort underpinning every aspect of Wikipedia. Equally important, volunteer editors have devised the policies and governance structures that frame the project, including dozens of unique policies across over two hundred language editions.\footnote{\url{https://en.wikipedia.org/wiki/Wikipedia:List_of_policies}}

Among these, the policy of Neutral Point of View (NPOV) stands as a cornerstone—it is, in fact, one of Wikipedia’s five foundational pillars.\footnote{\url{https://meta.wikimedia.org/wiki/Wikimedia_Foundation_Annual_Plan/2025-2026/Global_Trends/Common_global_standards_for_NPOV_policies/Analysis_of_Neutral_Point_of_View_Policies\_across\_Wikipedias\#General\_Context}} According to the English Wikipedia NPOV policy, editors are expected to ``strive for articles with an impartial tone that document and explain major points of view, giving due weight for their prominence.''\footnote{\url{https://en.wikipedia.org/wiki/Wikipedia:Five\_pillars}} While NPOV is central to Wikipedia’s mission and identity, surprisingly little rigorous scholarly attention has been devoted to examining the policy and its implementation in practice.

\subsection{What is the problem?}

One reason behind the scarcity of research on Wikipedia's NPOV policy may stem from the domain's complexity: studying NPOV requires in-depth knowledge both of Wikipedia’s inner workings and of the intricate ways in which neutrality is defined and operationalized within the project. In contrast, similarly complex domains—such as fairness and bias—have attracted robust research and interdisciplinary contributions.

\subsection{Why is the lack of research about NPOV important to Wikipedia?}

Continued progress and renewed vitality in Wikipedia rely, in part, on sustained, high-quality research spanning its more than 300 language editions. Given the centrality of NPOV and the global movement toward a more decentralized internet, rigorous study of neutrality on Wikipedia is more important than ever.\footnote{\url{https://www.noemamag.com/the-great-decentralization/}} The increasing pressures on Wikipedia and its editorial community to uphold neutrality only heighten the need for informed research and evidence-based improvements.

\subsection{Why is studying NPOV important to research?}

NPOV is one of the fundamental content policies on Wikipedia. The underlying processes, norms, and sources on which the Wikipedia community relies are of critical importance in interpreting Wikipedia articles, discussions, and community decisions. Wikipedia community members strive to achieve NPOV in editing, and often encounter challenges in doing so. Studying NPOV can reveal some of the thorniest issues involved in a project like Wikipedia. Insofar as researchers seek to evaluate Wikipedia in terms that align with the goals of the project and many of its contributors, NPOV represents an important set of dimensions for such evaluation. Finally, Wikipedia is so widely used as a source of trustworthy and accurate content that research into the content qualities as well as the processes that create its content can reveal more general insights into critical digital public infrastructure. 

\subsection{What has been tried so far to encourage and guide NPOV research?}

Some language communities within Wikipedia have created resources—such as NPOV FAQ,\footnote{\url{https://en.wikipedia.org/wiki/Wikipedia:Neutral\_point\_of\_view/FAQ}} NPOV tutorial,\footnote{\url{https://en.wikipedia.org/wiki/Wikipedia:NPOV\_tutorial}} and Controversial Articles Guidance\footnote{\url{https://en.wikipedia.org/wiki/Wikipedia:Controversial\_articles}} on English Wikipedia—to help editors interpret and apply the NPOV standard. However, no comparable resources exist to guide external researchers interested in conducting rigorous study of NPOV implementation and challenges. While Wikipedia maintains internal mechanisms, such as dedicated NPOV noticeboards,\footnote{For example, on English Wikipedia: \url{https://en.wikipedia.org/wiki/Wikipedia:Neutral\_point\_of\_view/Noticeboard}} for discussing NPOV concerns, these are intended for editors rather than for researchers, and focus primarily on internal deliberation processes rather than systematic, external analysis.

\subsection{What are we proposing to do?}

As stewards of Wikipedia, we see an opportunity and responsibility to support and equip researchers wishing to study NPOV. This white paper aims to lower the barriers to entry and foster high-quality inquiry by providing practical guidance for researching Wikipedia’s approach to neutrality. Our intention is that such research will not only advance scholarly understanding, but also produce actionable findings that benefit Wikipedia editors and the wider community.

This white paper does not seek to enumerate all of the ways in which researchers might or should study NPOV. While we, the authors, come from a variety of disciplines, we also have limitations. Most existing research focuses on English Wikipedia and our understanding of NPOV is largely rooted in that community's policies. At times we may refer to NPOV as if it's a singular policy, but each Wikipedia community may have their own local implementation. We often rely on English Wikipedia's policy as a referent given its influence and accessibility, but researchers should consider the local policies of language editions in which they ground their research. We also will be using terms like ``neutral'' or ``bias'' that may have very different meanings within different research disciplines or cultural contexts, but we chose to ground our terminology in that used by the Wikimedia communities as much as possible. These definitions are not always consistent\footnote{\url{https://meta.wikimedia.org/wiki/Wikimedia\_Foundation\_Annual\_Plan/2025-2026/Global\_Trends/Common\_global\_standards\_for\_NPOV\_policies}} and we welcome research that can help them evolve (see Section~\ref{sec:process-methods}).

The remainder of this paper is organized as follows: Section~\ref{sec:background} presents background and context to help researchers understand NPOV in practice; Section~\ref{sec:keyquest} outlines the core questions this paper seeks to address; Section~\ref{sec:recs} provides detailed answers.

\section{Understanding the context}
\label{sec:background}

\subsection{NPOV Definition and Principles}
The Neutral Point of View is one of Wikipedia’s foundational content policies. As codified in the English Wikipedia’s policy\footnote{\url{https://en.wikipedia.org/wiki/Wikipedia:Neutral\_point\_of\_view}} it reads:

``All encyclopedic content on Wikipedia must be written from a \textbf{neutral point of view} (\textbf{NPOV}), which means representing fairly, proportionately, and, as far as possible, without editorial bias, all the significant views that have been published by reliable sources on a topic.''

This policy, at the time of writing this paper, is formalized in 126 language editions.\footnote{As measured by the existence of a comparable title to the NPOV page on English Wikipedia.} It underpins Wikipedia’s goal of producing articles with an ``impartial tone that document and explain major points of view''\footnote{\url{https://en.wikipedia.org/wiki/Wikipedia:Five\_pillars}} with reference to the availability of reliable sources for the writing of an encyclopedia.

Several key concepts within the English Wikipedia policy and reflected similarly across many other language editions are worth elaborating:

\textbf{Proportionality (``Due Weight'')}: Coverage of each viewpoint should reflect its prominence in the available \textbf{reliable sources}.

\textbf{Freedom from editorial bias}: The written content should not reflect the personal opinions or preferences of Wikipedia editors and should be written using a tone that aims to be encyclopedic, that is neutral in tone.

\subsection{What NPOV Is; and What It Is Not}
NPOV is not the same as neutrality per se---instead, NPOV has particular definitions in Wikipedia. There are misconceptions associated with Wikipedia’s definitions of NPOV; these misconceptions are in some cases connected to wider discussions about what neutrality means, and in other cases very specific to Wikipedia, what the community is seeking to accomplish, and how it goes about its work. We will highlight some of the most important aspects of Wikipedia’s definitions of NPOV below and elaborate further.

\textbf{NPOV is about impartial summaries, not impartial content.} While the interpretation of NPOV has evolved over the years~\cite{steinsson2024rule}, generally NPOV requires that editors write an \textit{impartial summary} of significant viewpoints according to \textit{reliable sources.} This definition makes it difficult to assess the neutrality of the content itself. That is because the appropriate baseline that editors are aiming to achieve is the representation of viewpoints from reliable sources. This representation of viewpoints can be very difficult to establish and may be quite positive or quite negative in tone based on the sources themselves. Moreover, where the representation of viewpoints has clear majority and minority views, it will not result in giving equal weight to all viewpoints. This does not mean that every edit is itself neutral, but rather that the cumulative effect is towards a neutral representation.\footnote{e.g., see this editor essay: \url{https://en.wikipedia.org/wiki/Wikipedia:NPOV\_means\_neutral\_editing,\_not\_neutral\_content}}

\textbf{NPOV has two distinct facets: tone and due weight.} First, editors should avoid biased language when writing a Wikipedia article. This bias can manifest in a variety of ways such as stating opinion as fact\footnote{e.g., ``so-and-so is the greatest...''} or even stating facts as opinions.\footnote{e.g., ``Scientists believe that the Earth is round.''} Second, editors must balance sources by giving each viewpoint due weight, based on its representation among reliable sources and considering the topic of the article.\footnote{e.g., the belief that the Earth is flat is a fringe view and therefore should not be given undue weight or prominence in an article about the Earth though it is acceptable as a standalone article about the view given its long history.} The due weight component of NPOV is just as important as the first one, but often much harder to assess.

\textbf{Editing in pursuit of NPOV does not aim to resolve controversy but to reflect it.} For contentious topics, NPOV means describing the controversy (including criticism) rather than resolving it or adopting a ``middle ground.''

\textbf{NPOV can give due weight to biased sources that are reliable.} Wikipedia distinguishes between the terms ``bias'' and ``reliability'' when considering sources. A biased source---i.e. a source representing a particular viewpoint---is not \textit{automatically} unreliable – i.e. a source misrepresenting facts. For example, a government source may present only an official viewpoint and as a result can be biased. But that same government source is likely highly reliable (i.e. factually accurate) when looking to determine that government’s position on an issue. What matters for NPOV is the responsible and proportional representation of viewpoints from \textit{reliable} sources considering the context of the article, even if those sources are not ``neutral'' or ``unbiased.''

\textbf{NPOV is not determined based on a definitive list of acceptable or unacceptable sources.} Wikipedia does not maintain a pre-approved list of sources or list of sources that cannot be used under any circumstances. Decisions about a source’s reliability are made contextually—often for each specific use, and sometimes through community discussion or escalation to noticeboards such as the Reliable Sources Noticeboard. In English Wikipedia, the Perennial Sources list\footnote{\url{https://en.wikipedia.org/wiki/Wikipedia:Reliable\_sources/Perennial\_sources}} summarizes current consensus on frequently contested sources but does not result in inflexible rulings.\footnote{\url{https://en.wikipedia.org/wiki/Wikipedia:Reliable\_sources\#Context\_matters}}

\subsection{Dimensions and Factors Influencing NPOV}
There are multiple dimensions and factors influencing NPOV. We will share the most important ones below. 

\subsubsection{Language and community differences}
Wikipedia is not a monolith – it has over 300 different language editions, each with its own community. While different Wikipedia communities share key principles, their rules and the practical application of NPOV may differ by language. Below we share some of the factors contributing to these differences.

\textbf{Policy interpretation variance.} Although research has demonstrated qualitative evidence for key similarities in how Wikipedians apply NPOV in some Wikipedia languages~\cite{hickman2021understanding}, different Wikipedia language communities can have somewhat different interpretations of the NPOV policy.\footnote{\url{https://meta.wikimedia.org/wiki/Wikimedia\_Foundation\_Annual\_Plan/2025-2026/Global\_Trends/Common\_global\_standards\_for\_NPOV\_policies/Analysis\_of\_Neutral\_Point\_of\_View\_Policies\_across\_Wikipedias}}

\textbf{Policy enforcement variance.} Language edition communities enforce policy themselves, within the broader Wikimedia movement’s philosophy, leading to both diversity and (occasionally) inconsistency. As a result, different Wikipedias have adopted different approaches to NPOV~\cite{callahancrosslinguistic,massa2012manypedia}.

\textbf{Article topic variance.} Even within a given language edition, standards of NPOV and source reliability can differ by topic and subject area. For example, editors on English Wikipedia apply a more stringent set of source reliability standards to biomedical articles,\footnote{\url{https://en.wikipedia.org/wiki/Wikipedia:Identifying\_reliable\_sources\_(medicine)}} and may defer judgments for a specific topic to subcommunities specializing in that topic.

\textbf{Independence in coverage of a topic.} Different Wikipedia language communities are independent in how they choose to represent a given topic~\cite{johnson2022considerations}. While some communities use automated translation tools to assist with content creation,\footnote{\url{https://diff.wikimedia.org/2025/05/08/a-decade-of-consistent-improvements-to-the-content-translation-tool-yields-over-two-million-wikipedia-articles}} the number of articles in each language varies widely\footnote{\url{https://meta.wikimedia.org/wiki/List\_of\_Wikipedias}} and changes to content are often independent across language editions~\cite{hickman2021understanding}.

\textbf{Differences in resources.} There is substantial variation between language editions in the size of their editor community (e.g., number of active editors or users with extended rights) and the sources available to them. Some language communities have access to more open-access scientific and news resources across a range of viewpoints in their own language, while others have relatively few options in their own language.\footnote{The Wikipedia Library is one initiative to provide editors with access to publications but not all languages have localized support: \url{https://meta.wikimedia.org/wiki/The\_Wikipedia\_Library/Global}} These resource variations can affect the amount of time available and effort required by communities to develop and maintain content in different Wikipedia languages.

\subsubsection{Topic controversy and sources}
Whether an article is about a controversial topic or not affects the editors’ ability to implement NPOV. Sources for many historical events, scientific and mathematical topics, cultural and artistic movements, and biographies (among others) tend to agree on the major points about their subject matter and clearly identify the existing schools of thought where there are disagreements. In the case of controversial topics, and especially ongoing controversies where sufficient time has not passed to understand and organize sources and their positions holistically, maintaining NPOV can be difficult. This can be particularly challenging in areas where many sources disagree, where circumstances are changing quickly, or where editors do not agree on what sources they should use or how to cover them in a way that is fair and proportionate. Wikipedia has developed numerous guides for NPOV writing. English, for example, has an FAQ, a tutorial, and essays on specific topics including one focused on how to handle controversial topics as referenced in the Introduction. What this often means is that Wikipedians in the aggregate strive to describe, rather than resolve, a controversy by using encyclopedic tone. Further, Wikipedians do their best to create articles that reflect the relative number and significance of the sources available.

\subsubsection{The choice of the baseline}
\textbf{The baseline for NPOV is heavily influenced by reliable sources and referencing guidelines that every Wikipedia language adopts}, not by abstract, unattainable ideals of objectivity or by balancing all opinions equally. As a result, it is easy to oversimplify NPOV studies by studying individual components of Wikipedia in isolation and draw misleading conclusions about NPOV. High-quality content arises from the collaboration of editors with divergent viewpoints~\cite{shi2019wisdom,greenstein2021ideology} and editors must often balance biased (but reliable) sources to achieve high-quality content.\footnote{\url{https://en.wikipedia.org/wiki/Wikipedia:Neutral\_point\_of\_view\#Bias\_in\_sources}} Viewed in isolation, those editors, edits, and sources may appear biased, but together they may still achieve the goals of NPOV.

\subsubsection{The choice of time}
Wikipedia is a live project and is actively being edited. When you choose to study a portion of Wikipedia and what duration of time you choose to study it will have an impact on assessments of NPOV. Early in the history of the project, rules for sourcing, content, audience, and tone were still being developed. Research indicates that policies have stabilized in recent years~\cite{hwang2022rules}, so care is needed in retroactively examining policy compliance. 

Further, as a large and ever-evolving project, policy compliance also can shift over time. For example, consider current Wikipedia rules emphasizing reliable sourcing and frequent in-text citation. The number of sentences on English Wikipedia lacking citations has steadily dropped over the past 10 years. Correction speed has also changed over time. When a source is determined to be unreliable, the median time it is likely to remain in articles decreases by a factor of three~\cite{baigutanova2023longitudinal}. Over English Wikipedia's first ten years, bias related to US political topics steadily decreased, in particular as more articles were created and gaps were closed in the broader corpus of content~\cite{greenstein2012wikipedia}. Similar trends have been found when looking at articles that Wikipedians consider to be controversial~\cite{suresh2024s}. 

\subsection{The Role of Wikimedia Foundation}
\textbf{The Wikimedia Foundation serves as a steward of Wikipedia}, hosting technical infrastructure and supporting community self-governance. The WMF empowers local communities to develop and apply content policies such as NPOV, but \textit{does not} exercise day-to-day editorial control, taking direct action on Wikipedia only in rare cases due to legal or safety issues. Rather than acting as a conventional for-profit technology company, the WMF’s mission is to enable global access to free, community-developed educational content.

\section{Key Questions}
\label{sec:keyquest}
In this report, we seek to provide guidance on the following key questions:
\begin{enumerate}
    \item \textbf{How can I develop a good baseline when studying NPOV?} Neutrality cannot be evaluated in isolation – any research about bias on Wikipedia relies on a comparison between Wikipedia and some alternate representation of information.
    \item \textbf{How can I make rigorous assessments of Wikipedia's adherence to NPOV?} There are many different methods that have been used for measuring the neutrality of Wikipedia. Each has drawbacks but certain approaches are more robust and general open science principles can improve rigor and impact.
    \item \textbf{What are the best practices for communicating my research findings?} There are many potential audiences (Wikipedians themselves, the Wikimedia Foundation, the broader research community or world) for research on NPOV and it is important to appropriately contextualize your research to your audience.
\end{enumerate}

\section{Recommendations}
\label{sec:recs}
Having shared the context of NPOV on Wikipedia as well as the key questions we aim to answer in this paper, we now turn to providing recommendations corresponding to the key questions. In each section below, we divide the recommendations into one or more of the following categories: sources, content, editors, policy, and process.

\subsection{How can I develop a good baseline when studying NPOV?}
Assessing Wikipedia on NPOV, or any other dimension, requires comparison against a clear baseline or alternative perspective. We cannot overemphasize the importance of carefully choosing the baseline, understanding its shortcomings, and communicating transparently about it. Below we provide some guidance for the choice of baseline.

\subsubsection{Source baselines}
\textbf{Distinguish between bias in sources on Wikipedia vs. bias in sources outside of Wikipedia}. The references used by editors in Wikipedia articles underpin NPOV and the viewpoints represented within an article. As a result, Wikipedia can only cover information that reliable sources have already documented. Reliable sources themselves may be biased. Media bias pertains not just to how a particular topic is covered (``statement'' or ``framing'' bias) but also to which stories are considered worthy of coverage (also known as ``selection'' or ``gatekeeping'' bias)~\cite{castillo2025automated}. This can vary by what languages the story is covered in. Any assessment of source bias on Wikipedia should take into account whether this bias derives from gatekeeping bias by media organizations themselves or bias introduced by Wikipedians in how they select sources.\footnote{See \cite{saez2013social} for an example comparing news media and social media.} These two different pathways lead to entirely different assessments of Wikipedia as well as recommendations about how to address bias in Wikipedia.

\textbf{Avoid relying on any single source-rating system}. Many analyses of sources on Wikipedia rely on external source rating systems to characterize bias. These systems, however, do not necessarily measure the same concepts (e.g., ``bias'' most often, but sometimes ``credibility'' or ``reliability'') as Wikipedia editors, who draw an explicit distinction between ``bias'' and ``reliability''.\footnote{\url{https://en.wikipedia.org/wiki/Wikipedia:Reliable\_sources\#Biased\_or\_opinionated\_sources}} Even a relatively narrow notion of ``partisan bias'' turns out to be conceptually and empirically complex. Source-level bias scores can be misleading, as bias varies considerably at the individual story level for a given source~\cite{green2025curation}. In the US context for example, partisan bias in news sources is often the product of negative coverage of the ``out-party'' rather than positive advocacy on behalf of the ``in-party''~\cite{budak2016fair}, but Wikipedians have been fast to moderate negative coverage of politicians~\cite{kalla2015editorial}, which also raises the question of to what degree bias from sources translates into bias within content.

\subsubsection{Content baselines}
\textbf{Appropriately contextualize any comparisons to Wikipedia content}. There is no ``neutral'' corpus of knowledge against which to compare Wikipedia. In one sense, the most direct comparison cases for Wikipedia content are other online encyclopedias. However, many of these are ``forks''---diverging copies of Wikipedia---or else position themselves as Wikipedia alternatives that consciously adopt different policies and points of view (e.g., Baidu Baike,\footnote{\url{https://en.wikipedia.org/wiki/Baidu\_Baike}} the ruwiki fork,\footnote{\url{https://en.wikipedia.org/wiki/Ruwiki\_(Wikipedia\_fork)}} or Conservapedia\footnote{\url{https://en.wikipedia.org/wiki/Conservapedia}}, making empirical findings of differences potentially trivial. Many researchers have instead compared Wikipedia with Encyclopedia Britannica~\cite{krebs2024wisdom} as an example of an expert-written encyclopedia (Scholarpedia\footnote{\url{https://en.wikipedia.org/wiki/Scholarpedia}} might be another example). There are also many wikis with different notability criteria – e.g., Namuwiki\footnote{\url{https://en.wikipedia.org/wiki/Namuwiki}} or Fandom wikis.\footnote{\url{https://en.wikipedia.org/wiki/Fandom\_(website)}} Non-encyclopedic corpora such as newspapers, textbooks, or scientific journals also are potential points of comparison even if they are primary or secondary sources (as opposed to a tertiary source as Wikipedia is), though it may be harder to interpret any differences.

Choosing the right comparison depends on the focus of the study. Wikipedia’s coverage of recent events may be better compared against news sources, while Wikipedia’s coverage of specific historical or scientific topics may be better compared with textbooks or other expert-written encyclopedias. By contrast, coverage of culturally-specific topics may be best compared within Wikipedia but across language editions, e.g., German Wikipedia versus Japanese Wikipedia. We recommend that researchers be explicit as to \textit{why} they chose their points of comparison and what meaning they attribute to any differences.

\textbf{Consider differences in notability and style}. Comparisons must also account for variations in scope, genre, length, or other constraints. For example, differences in coverage of US politics between English Wikipedia and Britannica have been found to derive from Wikipedia articles being considerably longer on average than Britannica articles and covering a much broader range of topics (as opposed to differences in the specific tone used)~\cite{greenstein2018experts}. Making direct comparisons between matched pairs of articles on the same topics can help isolate differences~\cite{samoilenko2018don}, but may limit external validity if e.g., the matched set is systematically different from other articles. For example, focusing on current congressional representatives would mean that the articles fall under the Biographies of Living Persons policy\footnote{\url{https://en.wikipedia.org/wiki/Wikipedia:BLP} on English Wikipedia} on Wikipedia and would be subject to additional scrutiny by editors and perhaps editor or page restrictions as well.

This challenge is also faced by research that compares Wikipedia to scientific or news corpora. Given the differing goals and structure of these corpora, matching on topic for the purposes of comparing bias may be less meaningful unless articles on Wikipedia are compared directly to the sources they cite to understand differences in tone. Instead, research can explore what factors affect inclusion of a given source in Wikipedia such as open-access~\cite{dehdarirad2018type} or prominence~\cite{nielsen2007scientific,nicholson2021measuring}.

\subsubsection{Editor baselines}
\textbf{Be cautious about making normative statements about editor behavior}. Extensive empirical research has established that Wikipedia editing activity and approaches to editing are highly diverse~\cite{kim2016understanding,welser2011finding,yang2016edit} making normative editing behavior difficult to establish empirically. There are additional ethical and epistemological reasons to avoid defining a certain set of editors or editing behaviors as normative—and by implication, other editors or editing behaviors as ``aberrant''.\footnote{For a broader critique along these lines of NPOV as currently understood and practiced, see Menking et al.~\cite{menking2021wp}}

\textbf{Do not conflate editors contributing from a particular point-of-view with a violation of NPOV.} Wikipedia is structured to self-regulate towards NPOV as opposed to immediately achieving this with every edit to every article. As Wikipedia articles receive more edits, they have been shown to converge towards a more moderate tone~\cite{greenstein2018experts} as more-slanted articles attract editors with opposing views, the most biased contributors drop out, and others self-moderate their contributions. Greenstein et al.~\cite{greenstein2021ideology} and Shi et al.~\cite{shi2019wisdom} incorporate surveys of the political leanings of editors to US political topics and show the importance of having mixtures of editors with different points-of-view. They find that more diverse groups of editors engage in more constructive debates on talk pages and drive higher-quality content on Wikipedia, not just in political topics but also social and scientific ones.

\textbf{Assess edits and editor actions in context}. Edits examined in isolation may lead to incorrect conclusions as articles often change quite quickly. A problematic edit that only survives briefly (e.g., seconds) should be treated quite differently from problematic edits that survive long enough to be seen by many readers~\cite{priedhorsky2007creating}. Removal of content about a significant controversy may look like a violation of NPOV, but examination of other information may tell a completely different story. For example, the edit history might show that that content was added back in subsequent edits. Examination of administrative logs might reveal that the editor who added the content was topic-banned\footnote{\url{https://en.wikipedia.org/wiki/Wikipedia:Banning\_policy\#Bans\_apply\_to\_all\_editing,\_good\_or\_bad}} and so the removal was related to the editor and not the content. Examination of the edit summaries or talk page discussions associated with the edit might show the removal was part of moving the controversy to its own article in order to handle it more thoroughly, or the removal might have been temporary because the editors needed to identify more reliable sources to support the claims. Articles can be renamed, moved, or recategorized as part of maintenance activities in ways that make it difficult to connect current article state with past article state. Such context can be difficult to capture and trace, but studies that seek to characterize editing behaviors but do not account for context may reach inaccurate conclusions. Countermeasures against this kind of error include engaging in deep reading or other more qualitative examinations of articles and their context, engagement with community members about the research, and collaboration with experienced Wikipedia researchers.

\subsection{How can I make rigorous assessments of Wikipedia's adherence to NPOV?}
We start this section by providing some general guidance for how to do rigorous assessments of Wikipedia’s NPOV before moving to more specific recommendations related to source, content, editor, and process methods.

\textbf{Measure in multiple ways and report all of them}. For quantitative work in particular, there is no one "correct" way to measure adherence to NPOV. Each approach has its own benefits and drawbacks. Below we provide guidance for some reasonable ways to approach this challenge, but generally researchers should consider using multiple methods and triangulating the results to understand how sensitive their conclusions are to their method of analysis. 

\textbf{Adjust your approach to the level at which you evaluate content}. NPOV operates at different levels of content, which require different assessment and measurement approaches. At its most granular level, NPOV is relevant to individual sentences or paragraphs: 1) Is the word choice appropriate and the claim(s) backed by a reliable source? 2) Are all claims backed by reliable sources included? Simple classification models and approaches that look at a text excerpt in isolation or as compared to a single source are largely appropriate for this type of tone review. 

At the next level, NPOV applies to an article as a whole: are the individual claims being given weight due to their prominence in reliable sources? Here, more complex approaches that can extract claims, evaluate their prominence, and compare that to external sources would be necessary. Researchers may also need to make clear judgments about their points of comparison to determine whether they should only compare to the external sources selected by Wikipedia or should identify some other set of sources that they believe should be the point of comparison (and whether, after doing such a comparison, they believe that Wikipedia is missing sources that it should include). 

At the Wikipedia language edition level, NPOV applies to an entire topic area: are there major gaps in representation in terms of what articles exist and how thorough they are? Given the scale of content studied at this level, methods like those from the Knowledge Gaps literature~\cite{redi2020taxonomy} for tracking the prevalence of various sub-topics and their associated quality/framing or their trajectory in article-for-deletion discussions~\cite{tripodi2023ms} may capture some of the dynamics here.

\textbf{Use mixed-methods where possible}. NPOV is inherently contextual. Language describing an individual as a dictator may be biased in one context but an accurate reflection of reliable sources in a second. A source may be considered unreliable in one topic area but reliable in another. An editor may be pushing a POV in one topic area but a good-faith contributor in others. Impactful research will combine clear examples and qualitative analyses of NPOV with quantitative data to help in understanding how widespread of an issue these examples may be.

\subsubsection{Source methods}
\textbf{Clarify coverage when analyzing sources}. Analyzing source bias almost always means focusing on a particular subset of sources---e.g., academic or news media. Wikipedia articles combine a variety of sources though, which means care must be taken in generalizing findings from the subset of sources under study to articles or topic areas as whole. Yang and Colavizza~\cite{yang2024polarization} studied a corpus of 29M citations from English Wikipedia\footnote{This itself is not a complete dataset as it relies on usage of various cite templates in the wikitext, but not all references are generated in this way: \url{https://public-paws.wmcloud.org/55703823/HTML-dumps/references-wikitext-vs-html.ipynb}} of which 85\% have a URL. They were able to map 16.6\% of all the citations to a polarization score using the external Media Bias Monitor dataset and 10.4\% also could be mapped to a reliability rating via Media Bias Fact Check. Their resulting analyses of polarization and bias by topic area are valuable, but remain based on a small, non-representative fraction of the sources actually in use. Likewise, Baigutanova et al.~\cite{baigutanova2023comparative} studied English Wikipedia's Perennial Sources list and found that only 23\% of articles cited at least one source from this list, which includes a mixture of reliable and more questionable sources (with lower coverage on other language editions).

\textbf{Do not treat all citations as equal}. A challenge facing (quantitative) research around sources and neutrality is that not all citations to a source are equally important or reflect similar intent. Readers engage with citations in very unequal ways with more attention going to citations at the top of articles, in lower-quality articles, to open-access content, and when they relate to aspects of people's social lives~\cite{piccardi2020quantifying}.

The context in which a source is used also matters for considering its role. For example, at the time of writing this paper, English Wikipedia has established that Twitter (now X) is generally unreliable as a source,\footnote{\url{https://en.wikipedia.org/wiki/WP:RSPTWITTER}} but does allow usage if someone is describing themself. CNET is considered generally reliable only before October 2020.\footnote{\url{https://en.wikipedia.org/wiki/WP:CNET}} Forbes is considered generally reliable only when the author is a member of their staff (as opposed to a generic "contributors" byline)\footnote{\url{https://en.wikipedia.org/wiki/WP:FORBES}} HuffPost is considered reliable, but only in topics that are not about politics.\footnote{\url{https://en.wikipedia.org/wiki/WP:HUFF}} Other sources may be considered reliable but are known to exhibit bias\footnote{\url{https://en.wikipedia.org/wiki/Wikipedia:BIASED}} and so editors are expected to clearly attribute the source in the text and not give undue weight to the viewpoints expressed. Capturing this nuance is a difficult limitation facing most work. When reporting statistics about source usage and reliability, researchers can consider spot-checking individual examples to determine in which ways a source is being cited.

\subsubsection{Content methods}
\textbf{Be purposeful when choosing an assessment strategy and clearly describe what aspect of NPOV it captures}. NPOV covers both the balance of viewpoints by reliable sources and the tone used to express them. Despite NPOV being about both of these facets, most (quantitative) research focuses solely on the tone component. There are many more tools available for automatically assessing bias in tone and it can be done by focusing just on what is present in the article (as opposed to assessing what content might also be missing). Common assessment approaches are discussed below. The knowledge gaps literature---e.g., assessment of gender bias within Wikipedia~\cite{park2021multilingual,wagner2015s,redi2020taxonomy}---has substantive overlaps with the broader NPOV literature and can also be a source of inspiration.

\textbf{Use sentiment analysis only in tightly-controlled settings}. Research that uses sentiment analysis generally mixes together the balance of viewpoints and tone components of NPOV. Sentiment analysis does not usually distinguish between e.g., a sentence that describes a person's death in a neutral manner and one that describes a neutral concept but with negatively-charged language. It is important not to conflate differences in sentiment with differences in tone alone. Sentiment analysis can be a powerful tool but likely only with carefully-matched samples, such as comparing article text with the language used by the cited sources to describe the topic.

\textbf{NPOV classification models can be a good fit for Wikipedia}. A number of past studies have used language models fine-tuned to detect NPOV violations based on trace data from editors~\cite{suresh2024s,hube2018detecting}. These models have generally shown high performance and are a good fit for assessing the tone in Wikipedia articles. Care should be taken, however, in applying them in other contexts. Wikipedia has a very specific encyclopedic style\footnote{\url{https://en.wikipedia.org/wiki/Wikipedia:Manual\_of\_Style}} that may be different from the approach taken by other corpora. Without further evaluation, these models are likely not appropriate for comparing Wikipedia with external sources as any differences may reflect more stylistic differences as opposed to true differences in the tone. Instead, additional data more appropriate to the corpus might need to be collected~\cite{sinno2022political}.

\textbf{Slant indexes can be appropriate for specific topics}. A large body of research focused on political polarization has taken the approach of constructing an index of terms that are strongly associated with a particular viewpoint~\cite{greenstein2012wikipedia}. These terms are themselves not always emotionally-charged and so would not necessarily be picked up by methods such as sentiment analysis. While these methods can pick up on more nuanced forms of language bias (or due weight), they are highly topic-specific, often low-coverage, and much of their robustness depends on careful construction of the slant index. For instance, Greenstein and Zhu~\cite{greenstein2012wikipedia} found that 60\% of English Wikipedia articles related to US politics did not match any of their keywords.

\textbf{Exercise caution assuming that LLMs are effective at evaluating NPOV}. Applications of large language models (LLMs) and related systems to filter, label, classify, and analyze text corpora are growing rapidly~\cite{ziems2024can}. Adapting such approaches to evaluate Wikipedia content offers a promising avenue for scalable and multilingual analysis, but comes with distinct threats to validity and risks of bias~\cite{ashkinaze2024seeing}. Complicating matters, Wikipedia content is a near-ubiquitous source of training data for most commercial LLMs, raising questions about the stability and validity of inferences drawn from LLM-based analysis of Wikipedia~\cite{kaffeeevaluations}. Work in this domain is evolving rapidly, so rather than suggest best practices or definitive guidance, we note that existing literature indicates the need for caution and sensitivity to the various ways that LLM-based approaches can introduce distinct limitations or threats.

\textbf{Due weight should be assessed for both articles and topic areas}. When assessing whether Wikipedia is presenting an appropriate balance of all major viewpoints, this criteria should not just hold for individual articles but also for larger topic areas. For instance, individual articles may be well-balanced but if there is a bias towards articles about notable negative events for just one side of a controversial area but not the other, this can also introduce NPOV concerns. For example, research has found that topic-level bias (related to US politics on English Wikipedia) is largely a function of content gaps as opposed to the content of the existing articles~\cite{brown2011wikipedia,greenstein2012wikipedia} while there are mixed results on bias in article-deletion dynamics related to topics like the gender gap on Wikipedia~\cite{tripodi2023ms,worku2020exploring} Individual articles themselves should not grow too unwieldy sizes,\footnote{\url{https://en.wikipedia.org/wiki/Wikipedia:Article\_size}} so editors will often split out content into separate articles as well, further complicating measurements~\cite{lin2017problematizing}.

\subsubsection{Editor methods}
\textbf{Focus on aggregate editor behavior}. As discussed more in the next section, assessments of individual editor activity is largely the domain of Wikipedia moderators. It can be valuable, however, to measure individual editing patterns and motivations with a goal of understanding aggregate behavior and content evolution. Because most Wikipedians edit pseudonymously, most research must either use Wikipedia-specific features such as edit count or infer social traits such as perceived political ideology from content traces as opposed to explicit declarations by the editors~\cite{shi2019wisdom}. We discuss below the different strategies for understanding editor behavior as it relates to NPOV.

\textbf{Keep in mind that topical interest alone does not indicate stance}. While research has shown that many editors specialize in a particular political party if they edit political topics~\cite{agarwal2020wikipedia}, research that has also gathered the stated political ideology of a sample of editors~\cite{shi2019wisdom,neff2013jointly} has found that these topical preferences only have weak correlations with the editor's ideology. As Shi et al.~\cite{shi2019wisdom} further emphasized, the collaboration between editors with differing opinions appears to be an important part of the process to achieving neutrality and quality. The slant of editors attracted to a given article can also change over time as editors moderate each other~\cite{greenstein2021ideology}.\footnote{This work infers editor ideology based on the language that a given editor adds or removes to articles in their edits, which is promising but has not been validated with external preference data.} Altogether, this body of research suggests that care should be taken in inferring editor ideology based on the articles they edit.

\textbf{Be aware that some editors self-report interests but this is a narrow and biased sample}. Some editors include boxes on their Wikipedia user page that describe their interests, including political ideology. This community of editors looks quite different from the larger population of registered users, so care should be taken in contextualizing findings in terms of editors who self-proclaim their political affiliation rather than generalizing~\cite{neff2013jointly}.

\textbf{Consider surveys but remember that sample biases are likely}. Surveys are an opportunity to gather the most robust data---a semi-representative sample of editors along with data about their activity on-wiki---but difficult to do well. Shi et al.~\cite{shi2019wisdom} is the best example of this: they worked with the Wikimedia community to identify a strategy for sharing surveys with editors without spamming them and ultimately sampled 500 editors and received 118 responses. External surveys\footnote{e.g., \href{https://web.archive.org/web/20250525171222/https://d3nkl3psvxxpe9.cloudfront.net/documents/Wikipedia_poll_results.pdf}{YouGov Survey: Wikipedia (2023.}} are a simpler approach but Wikipedia editors are relatively rare within the general population\footnote{e.g., estimates of U.S. internet users that have ever edited Wikipedia are around 8\%~\cite{shaw2018pipeline}.} and it is difficult to infer much from the data because editors can differ greatly in their activity level and impact on content on Wikipedia.

\subsubsection{Process methods}
\label{sec:process-methods}
\textbf{Consider deep mixed-methods research to provide recommendations related to processes on Wikipedia}. There is a large body of research that has looked at the evolution of policy~\cite{steinsson2024rule}, made critiques of existing policy or rule-making~\cite{menking2021wp,unreliableguidelines}, and analyzed the outcomes of discussions and how consensus is reached on Wikipedia~\cite{tripodi2023ms,kaffee2023should,im2018deliberation}. Where this research is successful, it generally mixes qualitative and quantitative analyses and a deep understanding of Wikipedia by the researchers themselves through editing, attendance at community events, or a long history of research related to Wikipedia. Trace ethnography~\cite{geiger2011trace} has been particularly effective for describing how these processes play out.\footnote{e.g., an analysis of moderating politically-motivated editing: \url{https://cyber.fsi.stanford.edu/news/wikipedia-part-one}}

\textbf{Seek to measure trends and not moments in time}. Content is constantly evolving on Wikipedia and can take time to reach a stable consensus and update based on new information. Measuring adherence to NPOV at a moment in time, especially if it is in a rapidly-developing topic area, may lead to conclusions that would be different if a later snapshot were used or if the state of content were compared over a duration of time.

\subsection{What are the best practices for communicating my research findings?}
Communication is key to the impact of research. Effective and responsible communication requires learning more about where, when, what, how and to whom to communicate the research findings. In this section we provide recommendations to researchers for communicating their research for impact to improve Wikipedia. 

\subsubsection{Where/when to communicate your research?}

\textbf{Communicate your research early and often where feasible}. We highly encourage researchers to open a dialogue with the community by creating a project page on Meta-Wiki\footnote{\url{https://meta.wikimedia.org/wiki/Research:Projects}} as soon as they have a research proposal ready. We also encourage researchers to keep that page updated, at least as they pass certain research milestones. Creating a public project page allows others in the Wikimedia spaces, including Wikipedia researchers, to learn about your project. This level of visibility reduces the element of surprise and gives an opportunity to others to share their feedback or tips as the research progresses. 

\textbf{Choose the correct audience for your communications in the Wikimedia Movement}. For the large majority of NPOV related research findings, insights and recommendations, Wikipedia communities on the ground are an important and eager audience for communicating about research findings.\footnote{Where you can communicate such information with the Wikipedia communities for maximum impact depends on the Wikipedia language that you have studied and what you want to communicate. Check out \url{https://meta.wikimedia.org/wiki/Help:Contents} and \url{https://meta.wikimedia.org/wiki/Research:FAQ} to learn more.} These are individuals or groups who can internalize the findings and make a decision as to whether to make adjustments or changes on the ground in response to research. On the other hand, the Wikimedia Foundation should be the audience of research communication if research findings identify violations of applicable law or serious threats to safety.\footnote{For legal issues, please see the Foundation’s contact page at \url{https://wikimediafoundation.org/about/contact/}. For safety issues, please see the more detailed contact information about the Foundation’s Community Resilience and Sustainability team, which includes our trust \& safety work, at \url{https://meta.wikimedia.org/wiki/Wikimedia\_Foundation/Legal/Community\_Resilience\_and\_Sustainability/Trust\_and\_Safety}} Of course, it is important to submit research for peer review and disseminate your work with the broader Wikipedia research community as well.

\textbf{Share back with the Wikimedia research community}. There is a large and active community of Wikipedia researchers. Every year, Wikipedia is the subject of hundreds of research publications, with many made available via preprint or open access. There are several annual workshops dedicated to Wikipedia-relevant research as well as research tracks in Wikipedia community conferences, a Wikipedia research newsletter, email lists, and more.\footnote{\url{https://meta.wikimedia.org/wiki/Research:Resources}} Similar to Wikipedia itself, many of these groups and venues aim to welcome new contributions and participants. If you have done research on Wikipedia, consider submitting your work to share with these communities or reaching out to people in leadership roles to learn how you can get involved. In addition, we highly encourage researchers to make knowledge gained from research on Wikipedia freely available as described in the Wikimedia Foundation Open Access Policy.\footnote{\url{https://foundation.wikimedia.org/wiki/Policy:Wikimedia\_Foundation\_Open\_Access\_Policy}}

\textbf{Consider sharing specific content or policy recommendations directly with Wikipedia editors}. Start with on-wiki places that are mostly narrowly-scoped to benefit from your research. This demonstrates an effort to connect with editors where they do their work and also avoids flooding the larger editor community with too many notices about research. This could be as focused as the talk page for a specific article you analyzed. More commonly this might be a WikiProject,\footnote{See the list of WikiProjects across different Wikipedia languages at \url{https://www.wikidata.org/wiki/Q4234303}} which are topical or task-related groups of editors that can be found on many language editions of Wikipedia---e.g., if you analyzed neutrality in the context of medical knowledge on English Wikipedia, you might post to WikiProject Medicine's talk page.\footnote{\url{https://en.wikipedia.org/wiki/Wikipedia\_talk:WikiProject\_Medicine}} Editors in these spaces (or any Wikimedians you worked directly with) will be able to guide you as to whether there are other places in which you could share your work. Wikipedians themselves read and review research about Wikipedia, publish perspectives in an internal newsletter,\footnote{\url{https://meta.wikimedia.org/wiki/Research:Newsletter/}} and collect lists of research articles about Wikipedia. Finally, you may also consider in-person convenings such as local meet-ups\footnote{\url{https://en.wikipedia.org/wiki/Wikipedia:Meetup}} or submitting sessions to larger gatherings such as Wikimania.\footnote{\url{https://meta.wikimedia.org/wiki/List\_of\_Wikimedia\_Conferences\_and\_Events}}

\textbf{Share legal or safety concerns with the Wikimedia Foundation}. If your research intends to make recommendations to the WMF directly, you are encouraged to share it, including during early stages to understand the WMF’s perspective on what you plan to recommend if this would be useful.

Note that when WMF is asked to address situations where Wikipedia editors are the decision-makers, we will suggest that you share your research with the corresponding editor communities and public more directly first. 

The WMF is available if you have questions about Wikipedia research (including by issuing this paper) and depending on context may be able to offer you advice on the best way to share your research if you are uncertain how to do so.

\subsubsection{How to communicate your research for maximum impact on Wikipedia?}

\textbf{Communicate in ways that strengthen Wikipedia}. Wikipedia is a public resource that relies on the good faith and commitment of hundreds of thousands of volunteers as well as the trust of its readers. As a researcher studying NPOV on Wikipedia you have immense power when communicating the results of your research. With that power comes responsibility, both in terms of what you communicate and how your communication can be received. As a rule of thumb, we recommend that when communicating about your research you ask yourself the question “Will this communication make Wikipedia weaker or stronger?” Critiques are valued but ideally are paired with constructive recommendations, are replicable, leave space for feedback from Wikimedians, and do not overstate conclusions. 

\textbf{Set aside time for sustained investment of time and energy for impact}. Research can make important, substantive contributions to Wikipedia, but this often requires sustained investment by the researcher in explaining their work, engaging with the editing community as well as the wider community of Wikipedia researchers. Disseminating research to these audiences as described earlier is a good first step.

\textbf{Prioritize the humans behind Wikipedia and their motivations}. Wikipedia editors are regular people trying to write good articles for the benefit of the world. This is not to say critique should not be direct, but framing research around how to help make Wikipedia better, rather than threatening or embarrassing editors trying to work on difficult topics tends to lead to a much better reception and a much higher amount of work being done to address an issue. 

\subsubsection{What to communicate?}

\textbf{Follow best practices around open science, including sharing data and code}. There are many small but consequential choices that a researcher must make when studying Wikipedia. For example, failing to account for Wikipedia redirects can decrease the correlation between edits and pageviews by almost 20\%~\cite{hill2014consider}. Failing to correctly account for the behaviors of bots and other automated processes can result in misleading conclusions~\cite{johnson2016not}. Transparent and reproducible methods such as sharing data and code can improve the replicability of analyses and enable other researchers to verify the findings from your study.

\textbf{Be as specific as possible}. It is helpful in communicating research about NPOV content coverage to be specific about the particular tone, edits, or sources that are the subject of concern or critique. Saying that a particular topic, article, or discussion does not meet NPOV principles can often lack enough detail for an effective review, particularly for long articles or broad topics where people may simply not know where to start and no single person is likely to be able to effectively review everything. Better research communication will identify particular word choices that editors can check (e.g., quotes of problematic paragraphs) or even better will use Wikipedia’s article histories\footnote{\url{https://en.wikipedia.org/wiki/Help:How\_to\_read\_an\_article\_history}} to link to specific content changes or stable versions of articles with the text at issue in the research identified. 

\textbf{Offer clarity about recommendations}. There are variations in the NPOV dimensions as well as NPOV problem types, which may need different solutions. While it is not incumbent on researchers to recommend the exact content that should be on Wikipedia (and indeed, doing so is not generally advised) it is immensely helpful to distinguish between concerns about sources explaining why they should not be used, concerns about how material is presented (i.e. that sources are appropriate but Wikipedia may not accurately reflect them or be giving them correct due weight), and concerns that critical material is missing requiring more to be added to an article but not necessarily requiring deletion of what is already present. Similarly, if recommendations are about process or policy, researchers should consider applicable editorial standards and practices that already exist within the relevant language edition(s), topic area(s), and articles.

\textbf{Identify what needs work now}. It is helpful when presenting research to distinguish between problems that have been resolved and problems that remain active or that need attention at the time of research. Wikipedia articles can go through substantial changes over time (particularly as new sources become available) and being clear about what issues represent historical problems as opposed to open issues at the date and time that the research is published will help Wikipedia editors know what needs work. 

\textbf{Communicate insights about editors with additional care}. Your NPOV research may result in insights related to Wikipedia editors, their work, their actions or decisions. Privacy is a core value for many Wikipedians. We ask that researchers remember that editors are humans too: they can make mistakes, and they are volunteers working to improve the information available to everyone as part of a complex global project. We have developed dedicated guidance for research that has a focus on editors and we strongly recommend that you consult with the guidance prior to (conducting and) communicating your research findings about editors~\cite{asikin2025research}.
\section{Conclusion}
\label{sec:conclusion}
NPOV is a fundamental policy that requires constant interpretation and effort from Wikipedia contributors and is not a simple matter of an article meeting a predefined set of criteria. In this paper we provided guidance for researchers on how to approach the challenge of assessing adherence to such a nuanced policy and communicating those results. No research approach is perfect but we hope this guidance helps researchers in making reasonable choices and properly contextualizing and communicating their findings, including limitations, to help improve Wikipedia.
\section{Acknowledgments}
We thank the members of the Wikimedia Research community who provided extensive feedback to our first draft.\footnote{\url{https://meta.wikimedia.org/wiki/Research\_talk:Guidance\_for\_NPOV\_Research\_on\_Wikipedia/Draft}} That feedback has been valuable in helping us to understand how this guidance is perceived and which aspects to focus on clarifying and expanding. We further thank the members of the Wikimedia Foundation’s Research team, many of whom provided early inputs about existing literature in this space and patterns to address in this sort of guidance. Finally, we thank the many individuals who serve or support the NPOV Working Group.\footnote{\url{https://meta.wikimedia.org/wiki/Wikimedia\_Foundation\_Annual\_Plan/2025-2026/Global\_Trends/Common\_global\_standards\_for\_NPOV\_policies\#Working\_group}} This work is better thanks to the formal feedback we received from these groups as well as the many informal conversations that they and others have graciously held with us.

\bibliographystyle{plain}
\bibliography{npov_guidance}

\end{document}